\documentclass{optica-article}

\journal{opticajournal} 

\articletype{Research Article}

\usepackage{lineno}
\usepackage{verbatim}
\usepackage{subcaption}
\usepackage{geometry}

\usepackage{adjustbox}

\usepackage{textcomp}

\begin{document}

\title{Optical modeling of systematic uncertainties in detector polarization angles for the Atacama Cosmology Telescope}

\author{Colin C. Murphy\authormark{1}, Steve K. Choi\authormark{2,1,3 *}, Rahul Datta\authormark{4}, Mark J. Devlin\authormark{5}, Matthew Hasselfield\authormark{6}, Brian J. Koopman\authormark{7}, Jeff McMahon\authormark{4}, Sigurd Naess\authormark{8}, Michael D. Niemack\authormark{1, 3}, Lyman A. Page\authormark{9}, Suzanne T. Staggs\authormark{9}, Robert Thornton\authormark{10} and Edward J. Wollack\authormark{11}}

\address{\authormark{1}Cornell University, Department of Physics\\
\authormark{2}University of California, Riverside, Department of Physics and Astronomy\\
\authormark{3}Cornell University, Department of Astronomy\\
\authormark{4}The University of Chicago, Department of Astronomy and Astrophysics\\
\authormark{5}University of Pennsylvania, Department of Physics and Astronomy\\
\authormark{6}Flatiron Institute, Center for Computational Astrophysics\\
\authormark{7}Yale University, Department of Physics \\
\authormark{8}University of Oslo, Institute for theoretical astrophysics\\
\authormark{9}Princeton University, Department of Physics\\
\authormark{10}West Chester University, Department of Physics\\
\authormark{11}NASA/Goddard Space Flight Center
}

\email{\authormark{*}steve.choi@ucr.edu} 


\begin{abstract*} 
We present an estimate of the Atacama Cosmology Telescope (ACT) detector polarization angle systematic uncertainty from optics perturbation analysis using polarization-sensitive ray tracing in CODE V optical design software. Uncertainties in polarization angle calibration in CMB measurements can limit constraints on cosmic birefringence and other cosmological parameters sensitive to polarization leakage. Our framework estimates the angle calibration systematic uncertainties from possible displacements in lens positions and orientations, and anti-reflection coating (ARC) thicknesses and refractive indices. With millimeter displacements in lens positions and percent-level perturbations in ARC thicknesses and indices from design, we find the total systematic uncertainty for three ACT detector arrays operating between 90--220 GHz to be at the tenth of degree scale. Reduced lens position and orientation uncertainties from physical measurements could lead to a reduction in the systematic uncertainty estimated with the framework presented here. This optical modeling may inform polarization angle systematic uncertainties for current and future microwave polarimeters, such as the CCAT Observatory, Simons Observatory, and CMB-S4.

\end{abstract*}


\section{Introduction}
\subsection{Cosmic birefringence}
Cosmic birefringence refers to the rotation of linear polarization on cosmological scales from parity-violating physics \cite{komatsu_2022}. Measuring the signature of cosmic birefringence would require a high fidelity polarimeter and could be limited by intrinsic offsets and systematic uncertainties in the polarization angle of the instrument. In this article we present a new framework to estimate the detector polarization angle systematic uncertainties from a polarization sensitive ray tracing software.

Detection of cosmic birefringence, or the rotation of linearly polarized light as it propagates over cosmological distances through empty space, would be evidence for charge, parity, and time reversal symmetry (CPT) violating physics \cite{Lue_1999}. Proposed pseudoscalar fields produced by axion-like dark matter candidate particles that couple to electromagnetism would modify the dispersion relation for electromagnetic waves in vacuum \cite{Carroll_1998}. This modified dispersion relation would differ for left and right circularly-polarized light, inducing a rotation in the linear polarization angle of photons that grows linearly with propagation distance. The cross-correlation between cosmic microwave background (CMB) polarization in $E$ and $B$ modes is sensitive to cosmic birefringence or other parity-violating physics \cite{minami2019simultaneous}. In a universe with no parity-violating pseudoscalar fields, $C_l^{TB}$ and $C_l^{EB}$ terms in the CMB cross-correlation power spectra are zero because $Y_{lm}^E$ and $Y_{lm}^T$ have opposite parity to $Y_{lm}^B$. Cosmic birefringence mixes $E$ and $B$ polarization modes, which modifies $C_l^{TB}$ and $C_l^{EB}$ to $C_l^{\prime TB}$ and $C_l^{\prime EB}$in the following manner, where $\psi$ is the cosmic birefringence angle \cite{Abghari_2022}: 
\begin{equation}
C_l^{\prime TB} = -\sin (2 \psi ) C_{l}^{TE}
\end{equation} 
and
\begin{equation}
C_l^{\prime EB} = \frac{1}{2}\sin (4 \psi ) (C_l^{BB} - C_l^{EE}).
\end{equation}

Observation of a cosmic birefringence angle that is inconsistent with zero would have profound implications on fundamental physics. While constraints have been placed on the magnitude of cosmic birefringence from sources spanning the electromagnetic spectrum \cite{Marsh:2015xka, Hui_2017}, CMB experiments provide the tightest constraints on cosmic birefringence today \cite{Naess_2014, Choi_2020, Feng_2006, Kaufman_2014, Kaufman_2015, Xia_2008, Minami_2020}. 

Systematic uncertainties in the calibration of the polarization angles of polarimeters in CMB instruments produce a mixing of $E$ and $B$ polarization modes of the CMB that is degenerate with a detection of the cosmic birefringence angle. Historically, upper bounds on the cosmic birefringence angle have been limited by the statistical uncertainties of cross-correlation power spectra of CMB polarization. The most recent measurement of the cosmic birefringence angle $\psi$ by the Atacama Cosmology Telescope (ACT) experiment is consistent with zero ($\psi = -0.07^{\circ} \pm 0.09^{\circ}$\footnote{This uses opposite sign convention relative to some measurements \cite{Namikawa_2020}.}) \cite{Choi_2020}. The error bar is a weighted mean from a large heterogeneous sampling of $\psi$ with multiple detector arrays over several observing seasons with good overall consistency \cite{Choi_2020}. The ACT cryostat was removed from the telescope for replacements and installations of detector arrays then repositioned each year, suggesting a negligible level of systematic uncertainty. 
Nevertheless, improved statistical power in the forthcoming ACT data release (DR6) motivates an accurate estimate of the possible systematic uncertainty in the detector polarization angles. Additionally, several Galactic science measurements require precise polarization angle calibration \cite{Huffenberger_2020, Clark_2021, Vacher_2023, Cukierman_2023}. We present here a framework to estimate the polarization angle systematic uncertainty from optical modeling, demonstrated for the ACT experiment.

\subsection{The Atacama Cosmology Telescope and the AdvACT receiver}
\begin{figure}
     \centering
     \includegraphics[width=\textwidth]{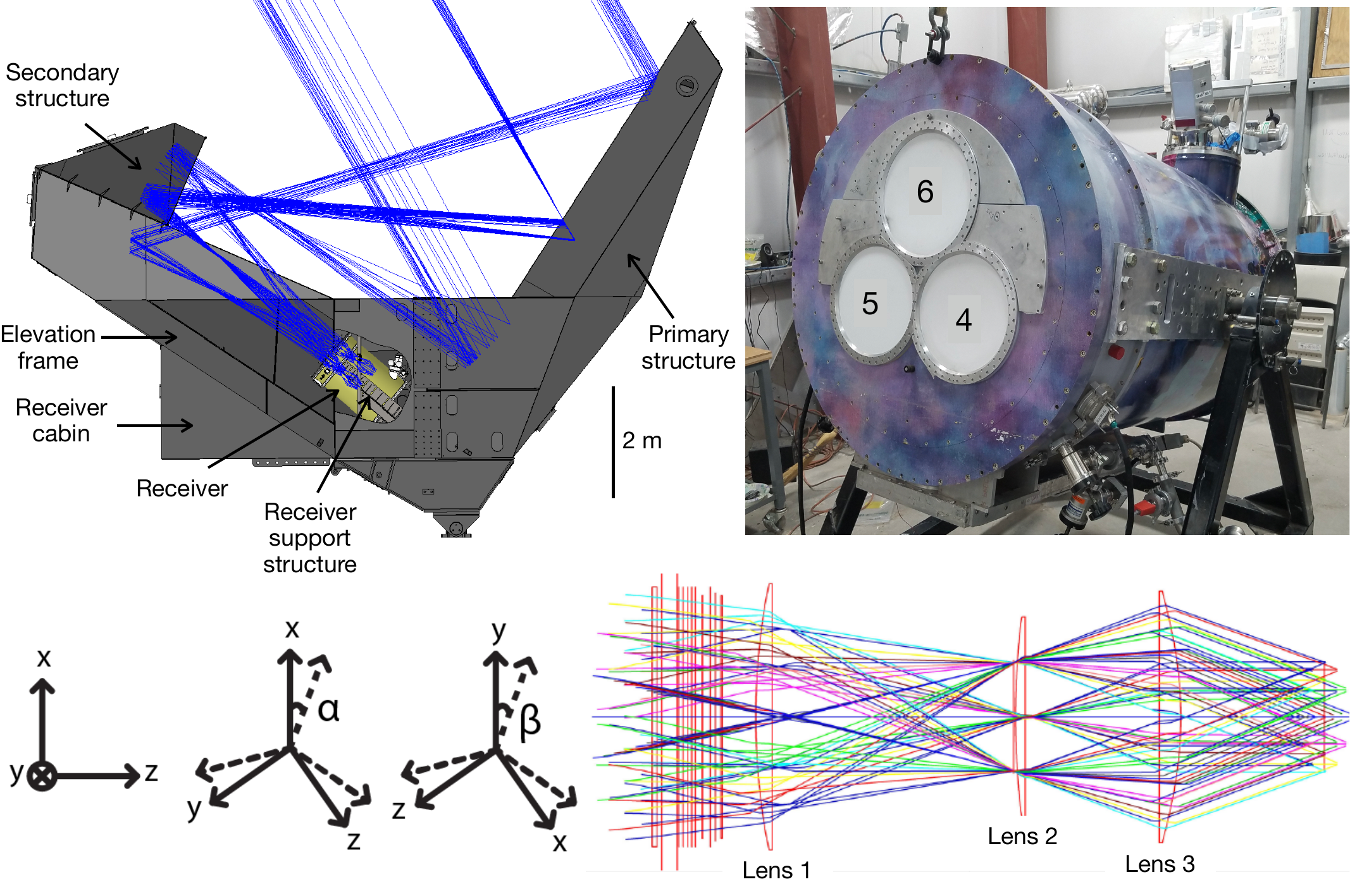}
     \caption{(Top left) Ray trace from the ACT Gregorian telescope mirrors to the AdvACT receiver. Figure adapted from \cite{Thornton_2016}. (Top right) The cryostat used for both the ACTPol and AdvACT detector arrays. In the AdvACT receiver, polarimeter array 4 (PA4) replaced ACTPol PA1, PA5 replaced PA2, and PA6 replaced PA3. Later, PA7 replaced PA6 for the last two years of observation. (Bottom) A top-down cross-section of the optical model of PA4 showing the three silicon lenses, the Lyot stop between lenses 1 and 2, and the $300 \textnormal{ K}$, $40\textnormal{ K}$, and 4 K low pass edge filters and IR blocker filter stacks \cite{Thornton_2016}. The $z$ axis is defined as the direction of propagation of the chief ray through the receiver optics. The light travels from left to right on the ray tracing figure shown at the bottom right. The $y$ axis is in the altitude direction, parallel to the axis of symmetry of the receiver, and the $x$ direction is in the azimuth direction, mutually orthogonal to the $y$ and $z$ directions in a right-handed manner. $\beta$ measures the angular displacement from the nominal $y$ axis, while $\alpha$ measures angular displacement from the nominal $x$ axis.}
     \label{fig:actpol_receiver}
\vspace{-0.2in}
\end{figure}

The Atacama Cosmology Telescope (ACT, 2007--2022) was a 6-m Gregorian telescope located at 5190-m elevation on Cerro Toco in the Chilean Andes \cite{Thornton_2016}. ACT's third and final receiver, Advanced ACTPol (AdvACT), included three dichroic arrays (PA4, PA5, and PA6) of dual-linear polarization-sensitive transition edge sensor (TES) bolometers, each housed in its own optics tube \cite{Henderson_2016, inproceedings, Choi_2018}. Fig.~\ref{fig:actpol_receiver} shows the telescope model, ray tracing, and a photo of the receiver. Each optics tube re-images a roughly 1$^{\circ}$ diameter area of the sky and contains three silicon lenses with silicon metamaterial anti-reflective coatings machined into them to account for their high index of refraction ($n$ = 3.4) \cite{Thornton_2016, Datta_2013}. With PA4--6, the AdvACT receiver was sensitive to radiation in three broadband frequency ranges centered at 98, 150, and 225 GHz \cite{Henderson_2016}. These frequency bands are labeled as f090, f150, and f220, and their bandpass shapes are found in \cite{Naess_2020}. PA4 is sensitive to radiation in the f150 and f220 bands, while PA5 and PA6 are sensitive to radiation in the f090 and f150 bands.  In 2020, PA6 was replaced with PA7, another dichroic array sensitive to 30 and 40 GHz \cite{Li_2021}. Preceding AdvACT, there were the ACTPol arrays, comprising two f150 arrays (PA1 and PA2) and one dichroic array sensitive to f090 and f150 bands (PA3) \cite{Thornton_2016}.

\subsection{The Simons Observatory, CCAT Observatory, and CMB-S4}

Next-generation large-aperture CMB observatories will map the CMB significantly faster due to larger detector counts filling larger diffraction-limited fields of view (FOV) \cite{Niemack:16, Parshley_2018}. The CCAT Collaboration's Fred Young Submillimeter Telescope (FYST) \cite{CCAT/2023}, the Large Aperture Telescope for the Simons Observatory (SOLAT) \cite{SO/2019,Gudmundsson_2021}, and the Chilean Large Aperture Telescope for CMB-S4 \cite{S4/2016,Gallardo_2022} share the same 6-m aperture Crossed-Dragone optical design. In Section~\ref{conclusion}, we briefly apply polarization-sensitive ray tracing to provide a preliminary estimate of the scale of systematic error in polarization angle measurements that can reasonably be expected for these observatories and compare it to ACT. 

\section{Method}
\subsection{Overview}
\begin{figure}[t!]
     \centering
     \includegraphics[width=\textwidth]{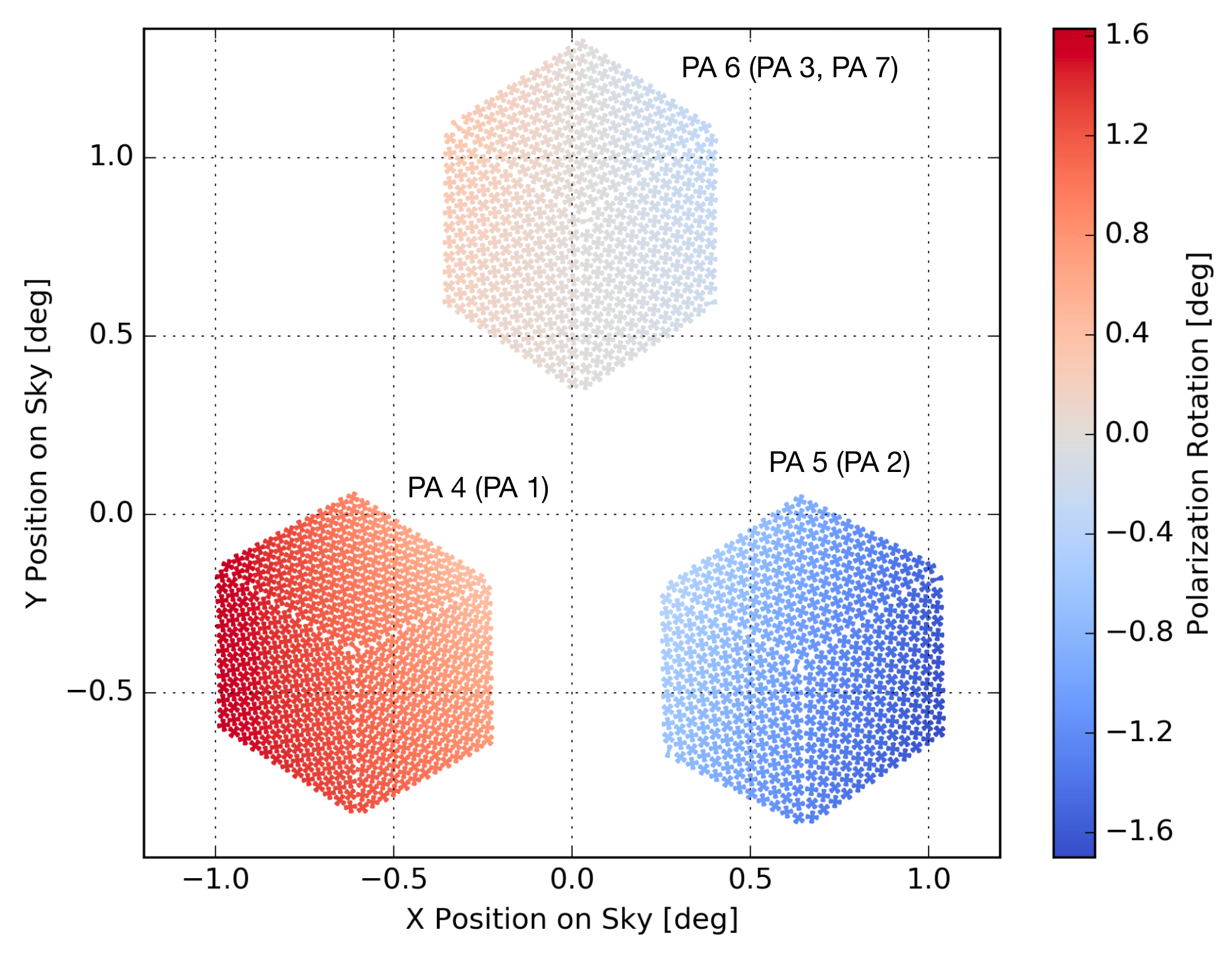}
     \caption{Detector sky positions and their nominal instrument polarization angles for the AdvACT detector arrays (labels in parentheses indicate other arrays used in the corresponding positions in different observing seasons). Note the axis of symmetry through the altitude direction and characteristic instrument polarization size of order 1$^{\circ}$. The three hexagonal groupings represent PA4--6, with PA6 in the top center, PA4 in the bottom left, and PA5 in the bottom right. PA4 is the high frequency array sensitive to 150 and 220 GHz. Figure from \cite{Koopman}.}
     \label{fig:advact_det_sky_coords}
\vspace{-0.2in}
\end{figure}

In this section we describe our method for estimating the polarization rotation. After passing through the ACT telescope mirrors and AdvACT refractive optics, light rays from the sky arrive at roughly localized points on the detector plane with a different polarization angle than they started with on the sky due to off-axis optics and refractive optics. The difference between the polarization angle for a given ray at the image relative to on the sky is henceforth referred to as the instrument polarization angle for that ray. The instrument polarization angles from the nominal receiver geometry have been calibrated (see Fig.~\ref{fig:advact_det_sky_coords}) across the field of view (FOV) at each detector location for PA1--6 \cite{Koopman}. Previous ACT constraints on the birefringence angle measurements (all consistent with zero) have corrected for this position dependent instrumental polarization, which can range up to $\sim$1$^{\circ}$ \cite{Namikawa_2020, Choi_2020}. While off-axis reflective optics can introduce polarization rotations \cite{Chu_1973,Jacobsen_1977}, any realistic mirror effects that can be simulated through physical and diffractive optics ray tracing are expected to be at least 10$\times$ smaller than the other refractive and alignment effects characterized here. Polarization (and far sidelobes) have also been studied for panel gaps in the mirrors and are noted to primarily induce a linearly polarized response \cite{Fluxa_2016}. However, polarization angle rotation, the subject of this study, does not naturally arise from these mechanisms.

The locations of the light rays on the image plane, along with their corresponding polarization angles, are sensitive to the orientation and locations of optical elements in the receiver and the thicknesses and indices of refraction of anti-reflective coatings (ARCs). Departures from the nominal optical model (e.g., the physical position of lens 3 being offset by 1 mm from the model) would result in instrument polarization angles that differ from the angle calibration derived from our simulations in Fig.~\ref{fig:advact_det_sky_coords}. 

In this paper, we estimate the systematic uncertainty in the instrument polarization angle calibration in \cite{Koopman} from optical simulations that include small perturbations in optical parameters. We assume that the systematic uncertainties in polarization angle calibrations corresponding to the uncertainty in different optical parameters are independent of each other. This is valid since each lens is fabricated separately and mounted to different physical portions of the optics tubes that are fabricated separately. The overall tilt in the optics tube assembly would affect the three lenses coherently, which is studied in this work through optics tube tilts and displacements. Therefore, we add them in quadrature to estimate the systematic uncertainty in linear polarization angle measurements for each array.

The lens and optics tube positions in each array are known to roughly 1 mm precision \cite{Datta_2013, Thornton_2016}. We adopt 1 $\textnormal{mm}$ precision as a baseline in this analysis. We perturb the positions of the three optics tubes and the three lenses in each of the optics tubes by $0$, $\pm 1$, and $\pm 2 \textnormal{ mm}$ along the $x$, $y$, and $z$ axis (see Fig.~\ref{fig:actpol_receiver}). We apply tilts of $0^{\circ}$, $\pm 0.5^{\circ}$, and $\pm 1^{\circ}$ to each of the three lenses in the $x$-axis (henceforth referred to as the $\alpha$ tilt) and in the $y$-axis (henceforth referred to as the $\beta$ tilt). For reference, a 0.4$^{\circ}$ tilt corresponds to displacing the ends of the $30 \textnormal{ cm}$ diameter lenses by 1 mm in opposite directions. The baseline uncertainties in the ARC thicknesses and indices of refraction are estimated to be 50 $\mu$m and 1\%, respectively \cite{Datta_2013}. Refractive elements not considered in this analysis include vacuum windows and metal-mesh filters. However, instrument polarization rotation is not expected from them, given the symmetry of the metal mesh structures \cite{Mackay_1989} and illumination at relatively small incidence angle and high focal ratio beams \cite{Ade_2006}.

\subsection{Detector coordinates}
\begin{figure}[t!]
     \centering
     \includegraphics[width=.9\textwidth]{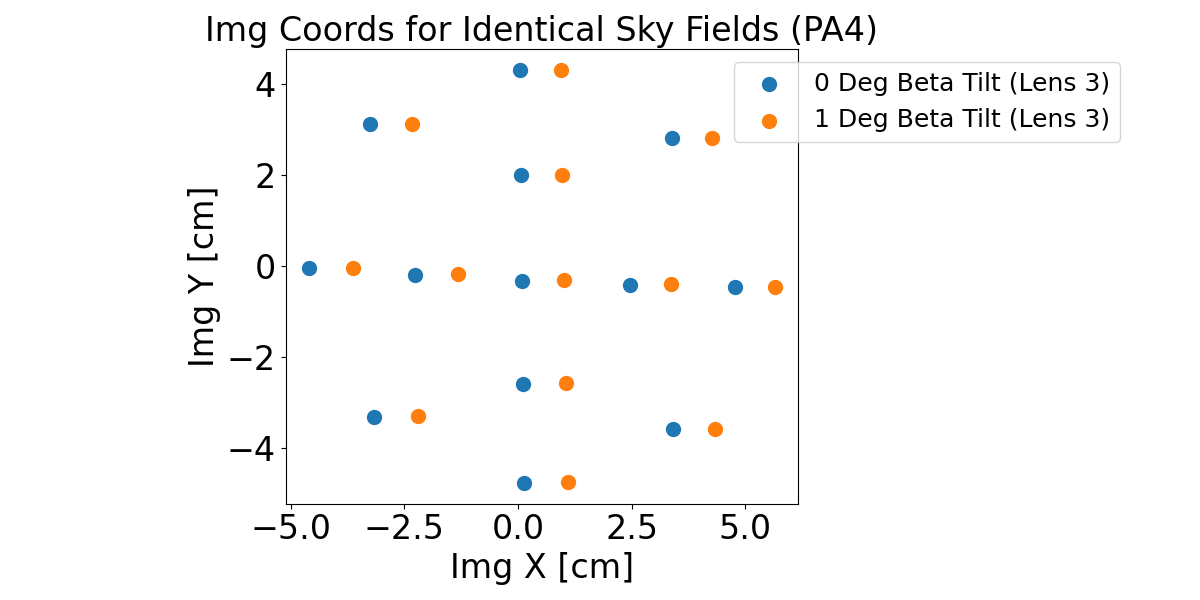}
     \caption{Change in locations of identical sky fields on the image plane due to a 1$^{\circ}$ $\beta$ tilt in lens 3 for one hexagonal array (PA4). Blue (orange) points show the original (modified) locations on the image plane before (after) the 1$^{\circ}$ tilt in lens 3 was applied. }
\label{fig:img_coords}
\vspace{-0.2in}
\end{figure}

The coordinates of each detector in the AdvACT receiver on the sky relative to the boresight have been calibrated through point source observations \cite{Choi_2020}. 
These source observations can account for and calibrate the small variations in the clocking of the optics tube during installation. Ray-tracing on the unperturbed optical models for each array was used to determine the location of detectors on the image plane (in mm) from their pointing-calibrated coordinates on the sky. The detector coordinates on the image plane are the true physical positions that are held fixed. Once a perturbation is introduced in the optical model of the array, the same physical detector locations on the image (i.e. the detector plane) correspond to different positions on the sky. Therefore, polarization angles in this framework corresponding to each detector in the array are calculated at the image coordinates of the detector. For example, Fig.~\ref{fig:img_coords} illustrates the level of offsets of the same fields on the image plane due to 1$^{\circ}$ $\beta$ tilt in lens 3 for PA4. 

\subsection{From a perturbed optical model to a systematic uncertainty in instrument polarization angle calibration}
\label{subsec:syst_uncert}
The level of instrumental polarization rotation in the ACT telescope is largely independent of the incident polarization angle \cite{Koopman}. In this framework, we consider the case where all of the incident sky fields are vertically polarized (i.e., all sky fields have a linear polarization angle of 90$^{\circ}$ relative to the horizontal).

Beginning with an unperturbed model of the telescope and array optics for an individual ACT array in CODE V (see Fig.~\ref{fig:actpol_receiver}), a small perturbation is introduced in one of the locations, orientations, or ARC parameters for an optical element in the array. The polarization definition for CODE V follows the Ludwig-1 \cite{Ludwig_1973}. The FOV is covered by 13 to 25 standardized fields (13 shown in Fig.~\ref{fig:img_coords}) that are propagated from the sky through the perturbed optics to the image plane using a polarization-sensitive ray tracing algorithm called \texttt{poldsp} (see Appendix~\ref{appen:macro} for more information about \texttt{poldsp}).  For some analyses, the most extreme fields are vignetted and only 13 are propagated through the system as in Fig.~\ref{fig:img_coords}. A cubic polynomial of the form $f(x,y) = Ax^3 + Bx^2y + Cxy^2 + Dy^3 + Ex^2 + Fxy + Gy^2 + Hx + Iy + J$ is fit to the image surface $x$ and $y$ coordinates of each of the fields and their polarization angles. This fit is then used to interpolate the polarization angles at each of the image $x$ and $y$ detector coordinates for the array. If the physical optics differ from the modeled optics by some perturbation, it would constitute a systematic uncertainty in the detector polarization angle calibration. The size of this systematic uncertainty would be set by the difference between the instrument polarization angle measured at a given detector location due to the optical perturbation and the nominal, unperturbed instrument polarization angle calibrations in \cite{Koopman}. For a given perturbation, we then average this systematic uncertainty at each detector location across the FOV to obtain the overall systematic uncertainty in polarization calibration for each array. 

Taking into account the detector sensitivities when weighting the FOV-average systematic uncertainty in polarization angle led to negligible changes compared to unweighted averages. Hence, array systematic uncertainty statistics reported in forthcoming sections are not weighted by detector sensitivities. 

We estimate the FOV-averaged (over detector positions) systematic uncertainty in instrument polarization angle calibration for four steps of perturbations at $\pm 1\times$ the nominal offsets in each of the optical parameters (e.g., displacing the lens position by $\pm$1 mm). These systematic uncertainties for each perturbed version of the optics were fit to the value of perturbation. The slope of the resulting line estimates the change in  instrument polarization angle as a function of the perturbation of the optical parameter. The uncertainty in the slope comes from the spread in the systematic uncertainty in instrument polarization angle over each detector array.

The overall array uncertainty is calculated from adding in quadrature the uncertainties due to perturbations of each of the three lenses in the array and the optics tube along $x$, $y$, $z$, $\alpha$, and $\beta$. In the equation below, $i=0$ refers to perturbations of the optics tube itself, and $i = 1, 2, 3$ refer to perturbations of the first, second, and third lens respectively:
\begin{align}
    \delta_{\textnormal{array}}^2 = & \sum_{i=0}^3 \left(
    \delta_{i,x}^2 + \delta_{i,y}^2 + \delta_{i,z}^2 +
    \delta_{i, \alpha}^2 + \delta_{i, \beta}^2 \right)
\end{align}

The standard deviations of the systematic uncertainties across the detector positions are also added in quadrature to indicate an uncertainty in the total systematic uncertainty, as indicated below: 
\begin{align}
    \sigma_{\textnormal{array}}^2 = & \sum_{i=0}^3 \left(
    \sigma_{i,x}^2 + \sigma_{i,y}^2 + \sigma_{i,z}^2 +
    \sigma_{i, \alpha}^2 + \sigma_{i, \beta}^2 \right)
\end{align}

Table 1 shows $\delta \pm \sigma$ for each of PA4, PA5, and PA6. Anti-reflective coatings on the lenses are modeled as three-layer coatings, where each layer has a different thickness and index of refraction to roughly match the measured frequency-dependent reflectance across the bands of interest. Using this approach, we find that systematic uncertainties from small perturbations of the anti-reflective coatings are negligible and omitted. We note that this model for the anti-reflective coatings does not take into account its subwavelength metamaterial nature \cite{Datta_2013}. As the wavelength approaches the size of the metamaterial structures (at high frequencies), the reflection properties deviate from the uniform index and thickness model used here \cite{Datta_2013}. Tilts on curved surfaces and uncertainties in the metamaterial structures could also impact the polarization response, but the challenge of modeling those effects accurately will require more advanced physical optics software and is beyond the scope of this work.

\subsection{Frequency dependence of instrument polarization}
\begin{figure}
     \centering
     \includegraphics[width=.9\textwidth]{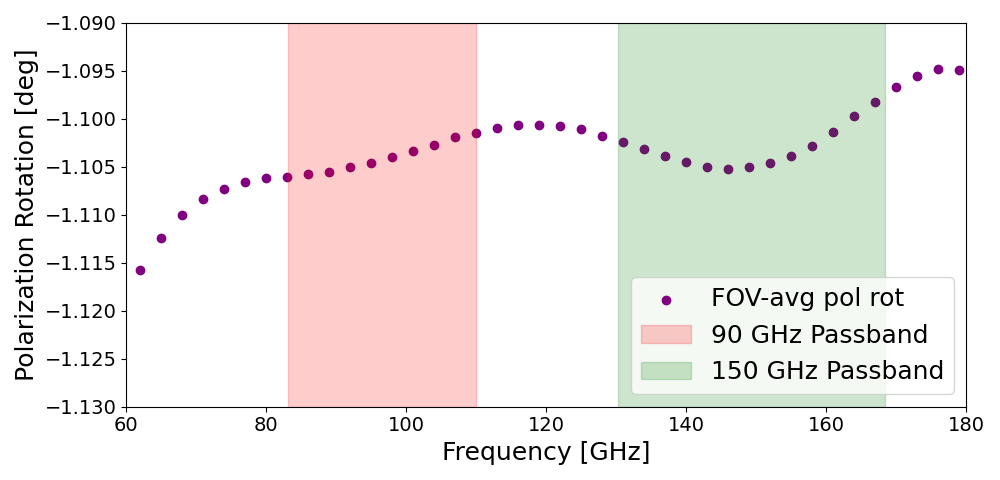}
     \caption{Frequency dependence of the array-average instrument polarization angle for PA5. Instrument polarization varies by less than 0.01$^{\circ}$ within the frequency bands of interest.}
     \label{fig:frequency_depedence}
\vspace{-0.2in}
\end{figure}

We perform simulations using incident light with the center of the band frequency for the lower of the two frequency bands in each array, as in Fig.~\ref{fig:frequency_depedence}. We assess the frequency dependence of the instrument polarization angle for PA5 between 50 and 200 GHz as a representative example. In particular, we consider the variation in FOV-averaged polarization angle in the two 20 GHz-wide frequency bands of interest, f090 and f150. We find that the instrument polarization angle does not depend on the frequency of incident light in our bands of interest at the 0.01$^{\circ}$ level, which is smaller than current measurement uncertainties. 

\section{Results}\label{results}
\begin{figure}[t]
     \centering
     \begin{subfigure}[b]{0.8\textwidth}
         \centering
         \includegraphics[width=\textwidth]{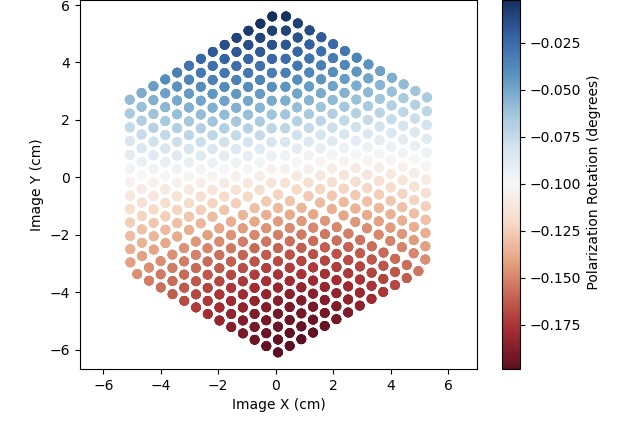}
     \end{subfigure}

        \caption{Change in instrument polarization  from nominal design across the detector plane for a $1.0^{\circ}$ $\beta$ tilt of the optics tube for PA4.}
        \label{fig:pa3 lens 1 beta tilt}

\end{figure}

\begin{figure}[t]
     \centering
     \begin{subfigure}[b]{0.3\textwidth}
         \centering
         \includegraphics[width=\textwidth]{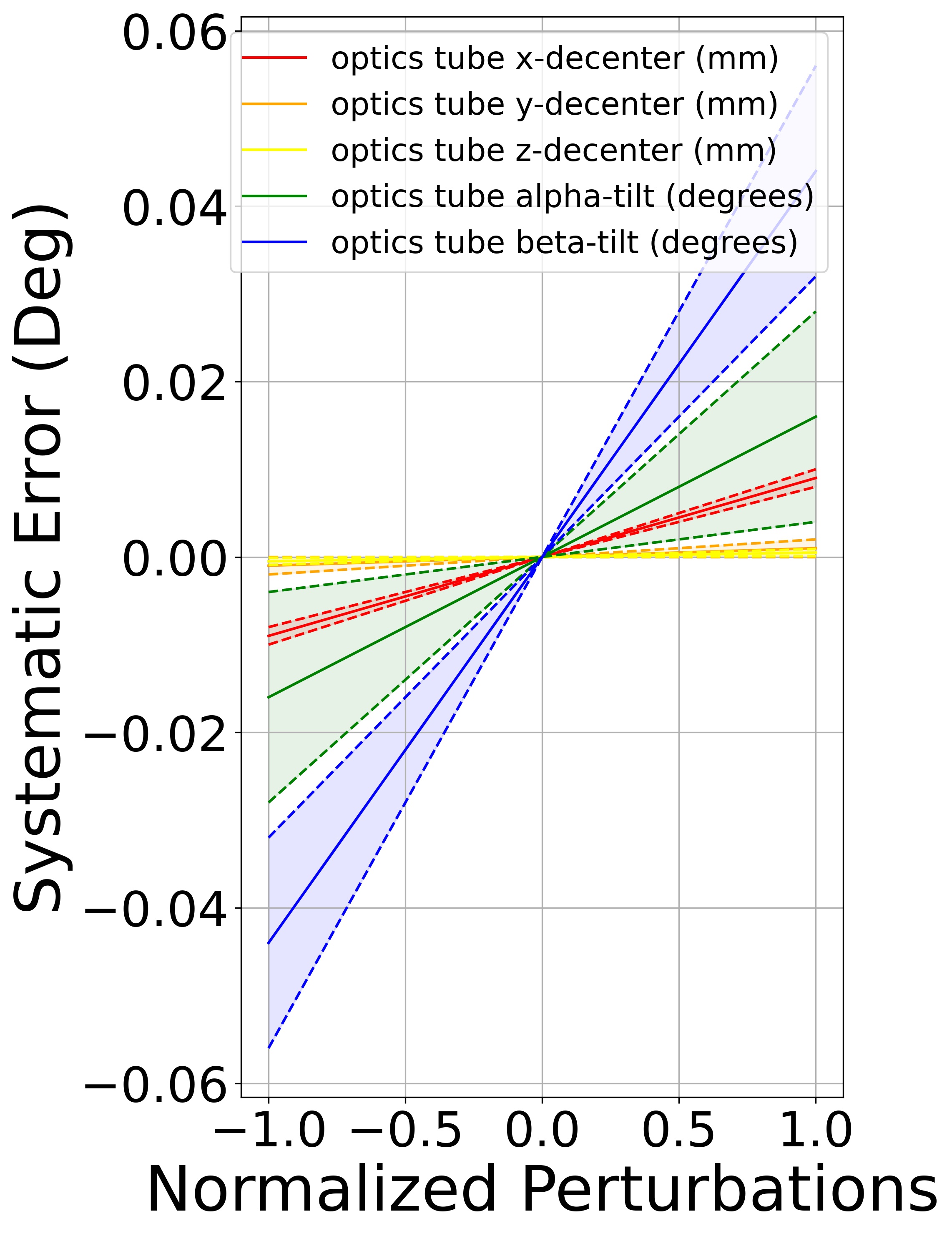}
         \vspace{-0.25in}         
         \caption{PA4 Optics Tube}
         \label{fig:pa4 optics tube}
     \end{subfigure}
     \begin{subfigure}[b]{0.3\textwidth}
         \centering
         \includegraphics[width=\textwidth]{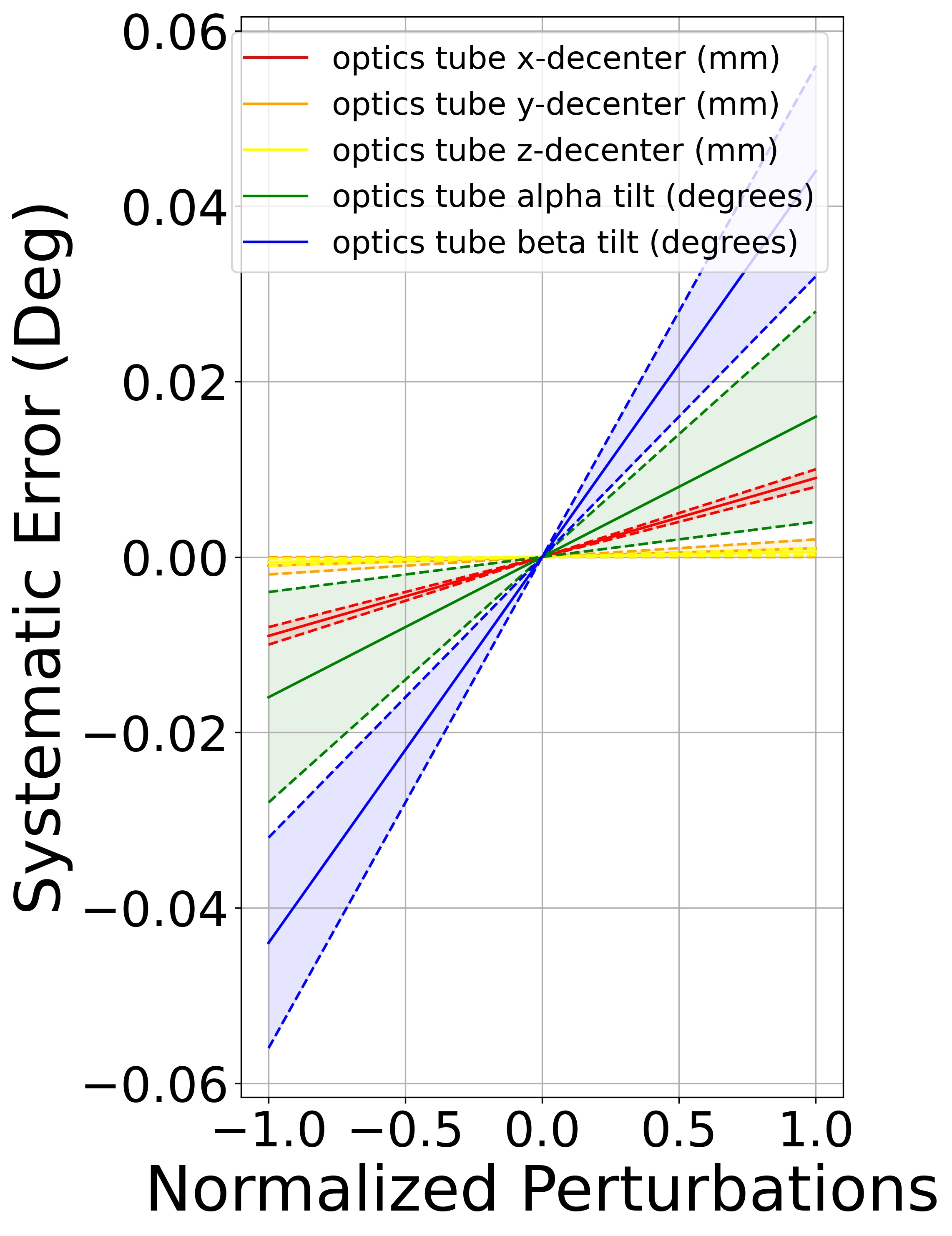}
         \vspace{-0.25in}
         \caption{PA5 Optics Tube}
         \label{fig:pa5 optics tube}
     \end{subfigure}
     \begin{subfigure}[b]{0.3\textwidth}
         \centering
         \includegraphics[width=\textwidth]{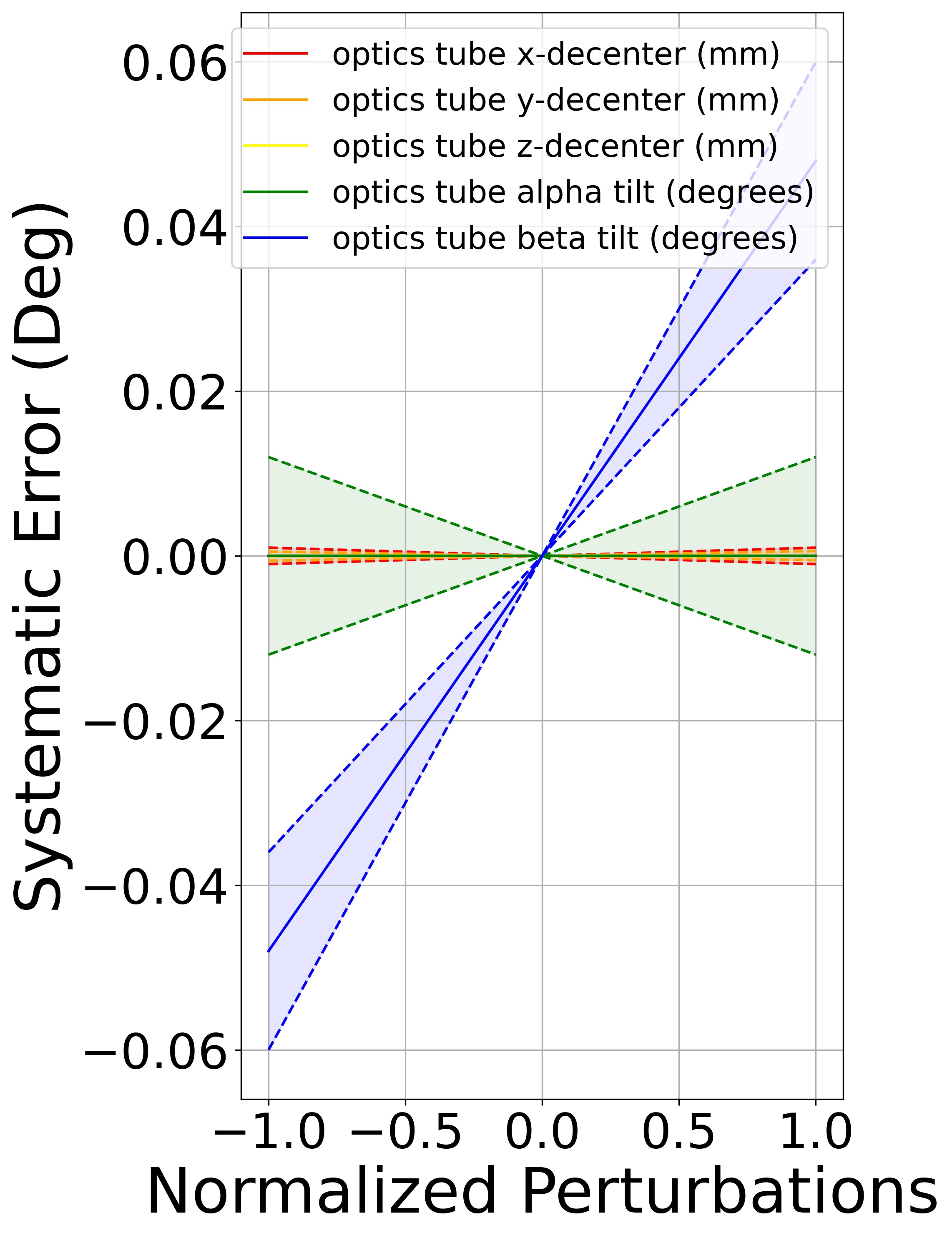}
         \vspace{-0.25in}
         \caption{PA6 Optics Tube}
         \label{fig:pa6 optics tube}
     \end{subfigure}
     
      \caption{Array-average systematic uncertainties in instrument polarization angles for each optics tube in PA4--6; dotted lines represent a single standard deviation of the slopes of each of the lines, corresponding to the spread of the systematic uncertainty in instrument polarization angles across the detectors. Perturbations are of $\pm1$ mm in $x$, $y$, and $z$, and $\pm$0.4$^{\circ}$ in $\alpha$ and $\beta$; these perturbations are normalized to enable comparison (i.e., in $x$, $y$, and $z$, an offset of $\pm$1 mm corresponds to a normalized perturbation of $\pm$1, and a tilt of $\pm$0.4$^{\circ}$ in $\alpha$ or $\beta$ corresponds to a normalized perturbation of $\pm$1). Perturbations of ARCs are not shown, as they are negligible.}
      \label{fig:FOV-average}
\vspace{-0.2in}
\end{figure}

\begin{figure}[ht!]
     \centering
     \begin{subfigure}[b]{0.3\textwidth}
         \centering
         \includegraphics[width=\textwidth]{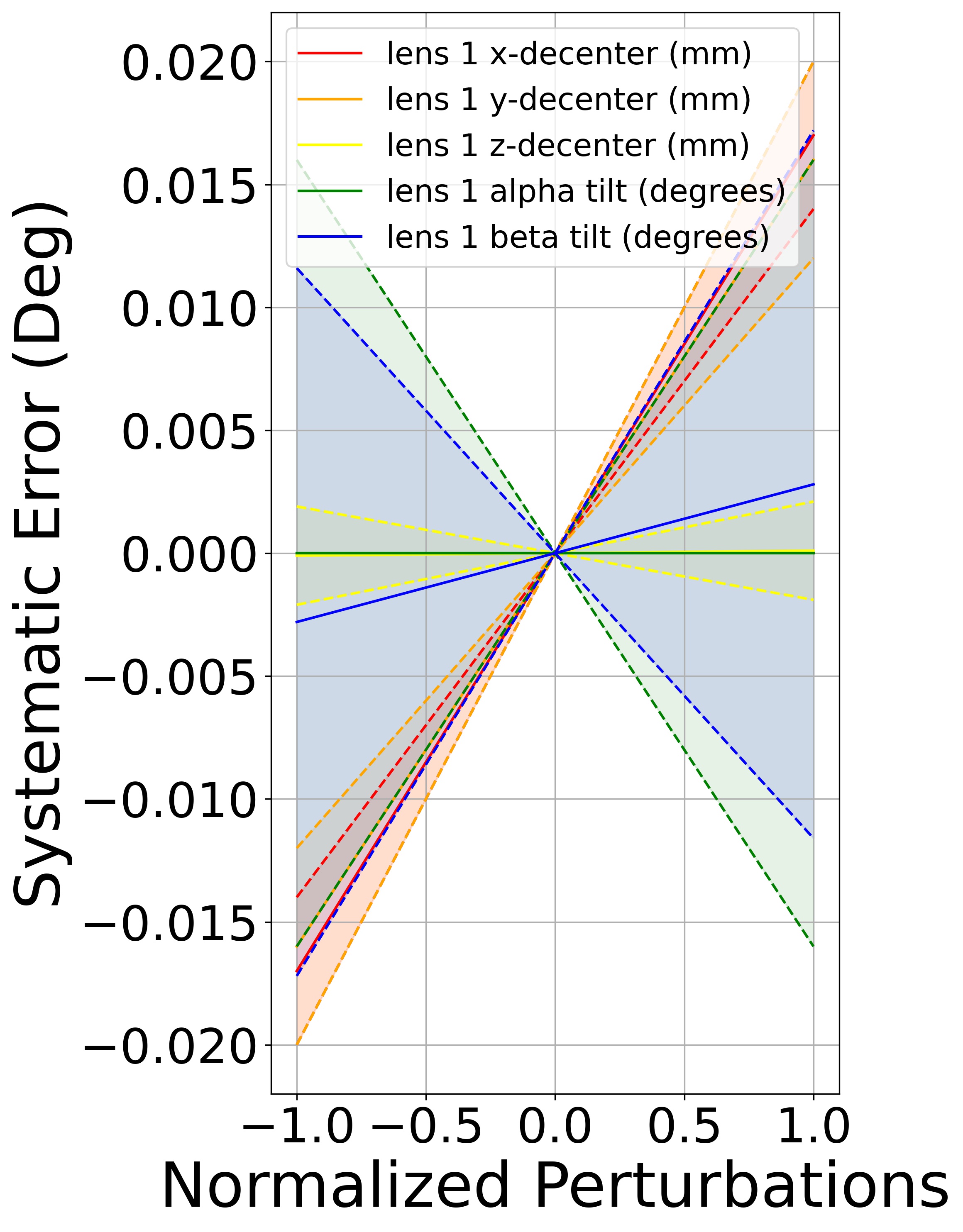}
         \vspace{-0.25in}
         \caption{PA4 Lens 1}
         \vspace{0.125in}
     \end{subfigure}
     \begin{subfigure}[b]{0.3\textwidth}
         \centering
         \includegraphics[width=\textwidth]{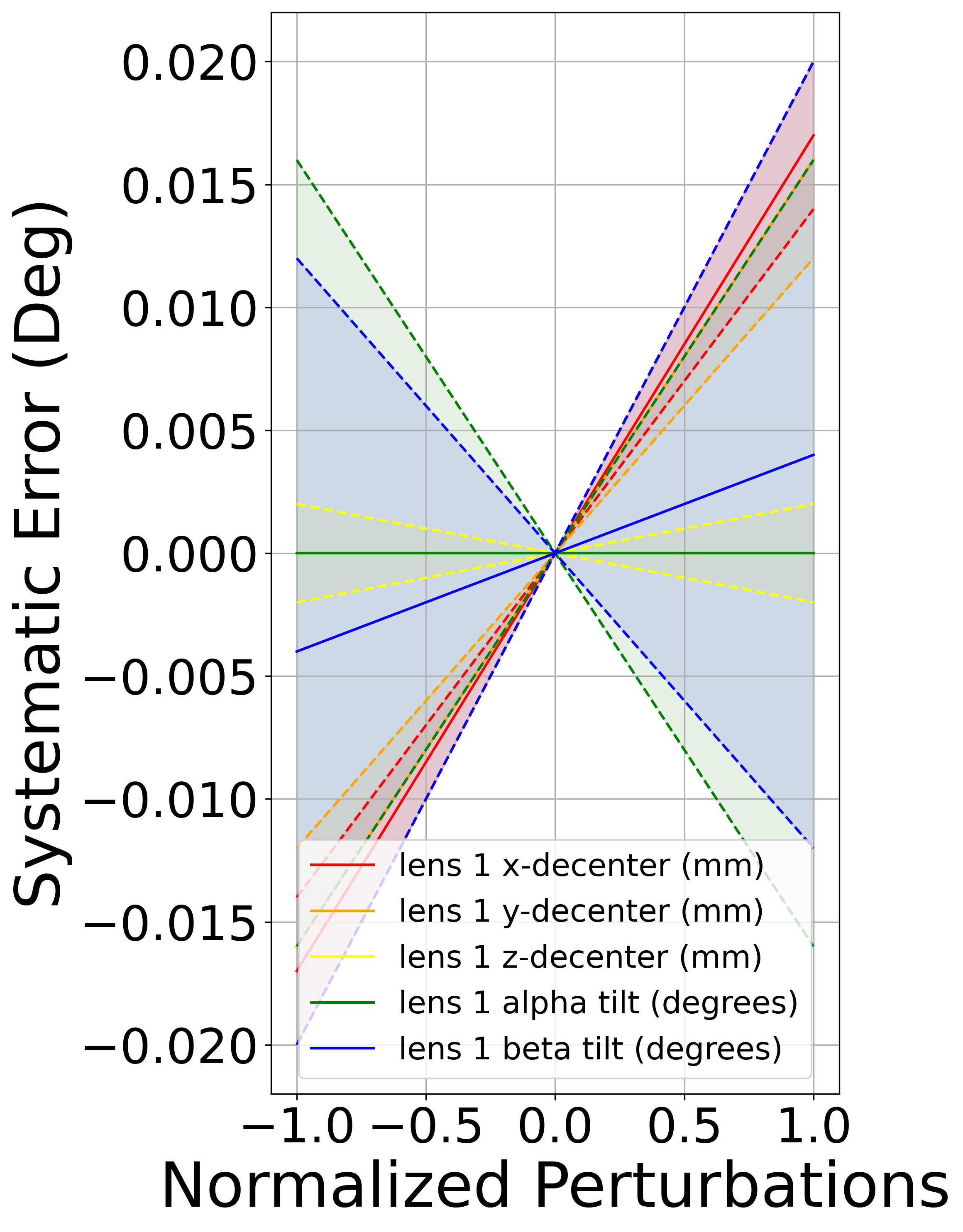}
         \vspace{-0.25in}
         \caption{PA5 Lens 1}
         \vspace{0.125in}
     \end{subfigure}
     \begin{subfigure}[b]{0.3\textwidth}
         \centering
         \includegraphics[width=\textwidth]{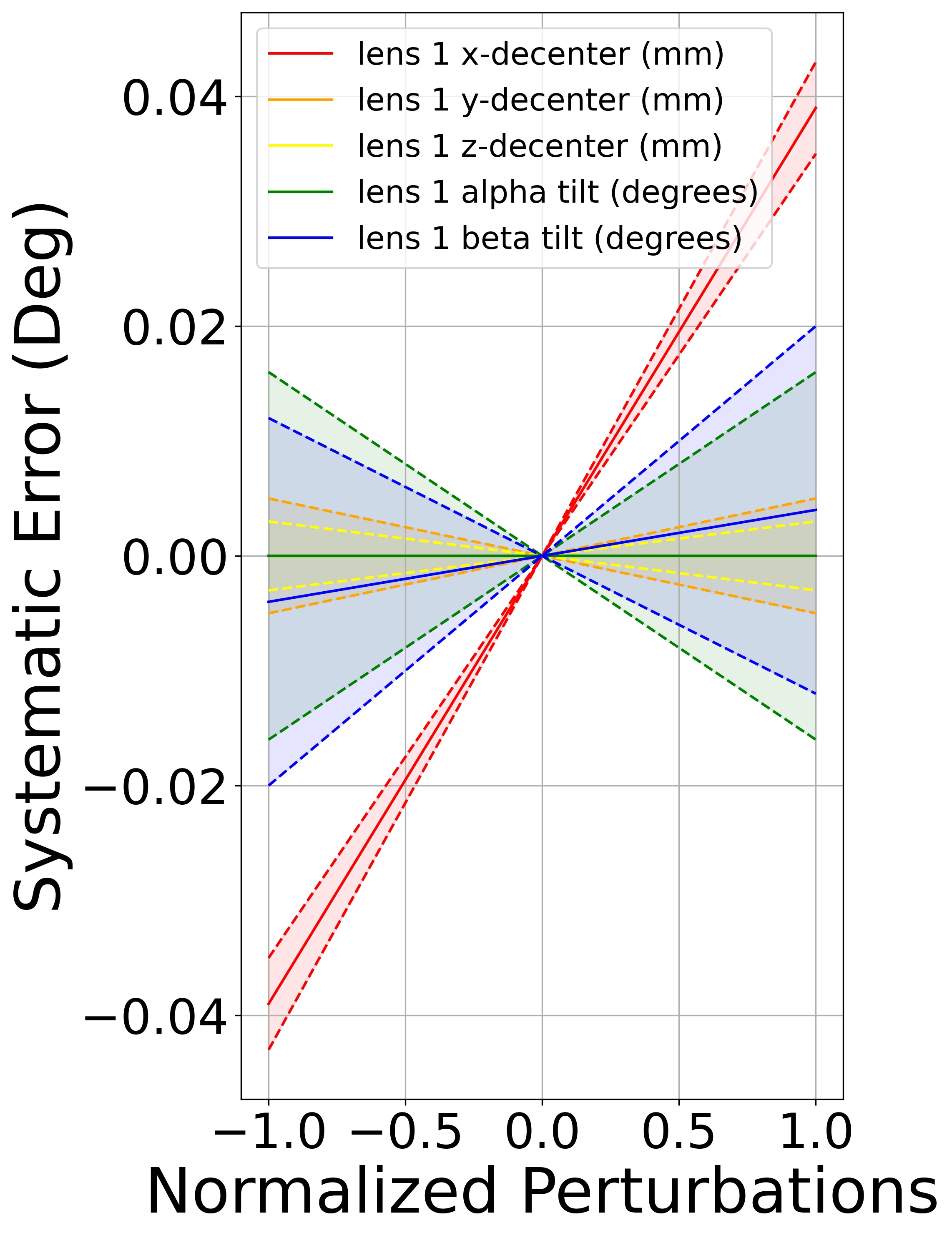}
         \vspace{-0.25in}
         \caption{PA6 Lens 1}
         \vspace{0.125in}
     \end{subfigure}
     \begin{subfigure}[b]{0.3\textwidth}
         \centering
         \includegraphics[width=\textwidth]{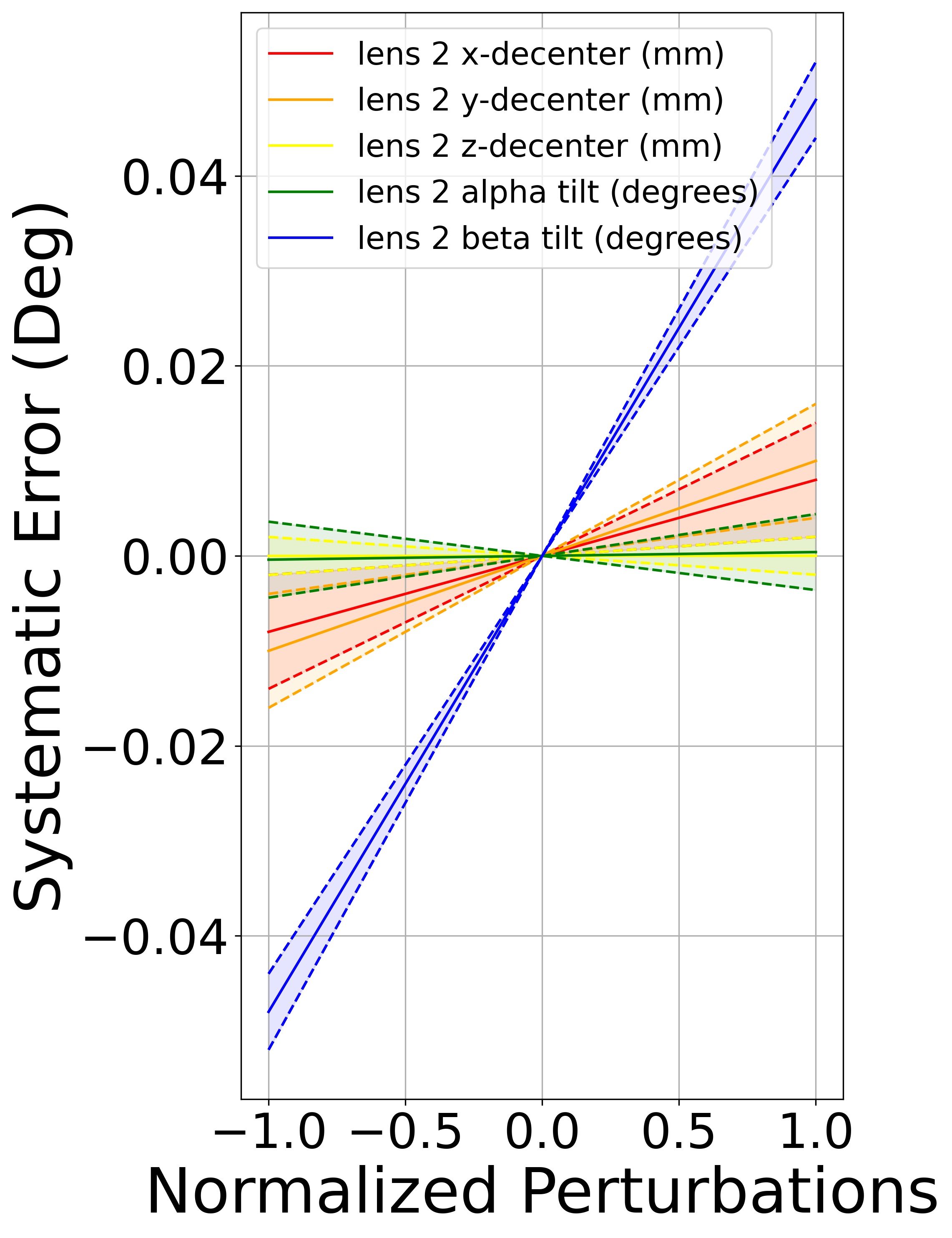}
         \vspace{-0.25in}
         \caption{PA4 Lens 2}
         \vspace{0.125in}

     \end{subfigure}
     \begin{subfigure}[b]{0.3\textwidth}
         \centering
         \includegraphics[width=\textwidth]{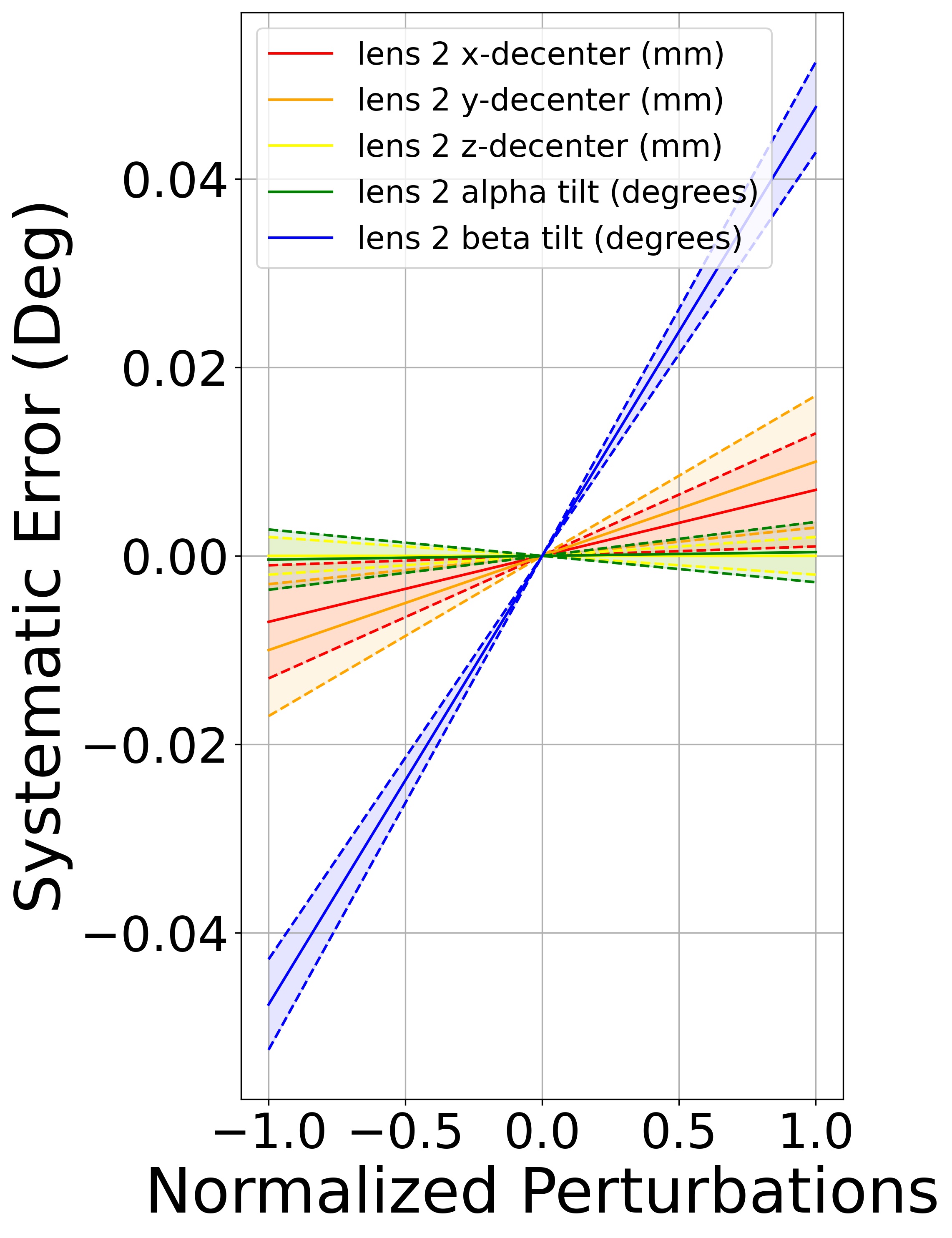}
         \vspace{-0.25in}
         \caption{PA5 Lens 2}
         \vspace{0.125in}
     \end{subfigure}
     \begin{subfigure}[b]{0.3\textwidth}
         \centering
         \includegraphics[width=\textwidth]{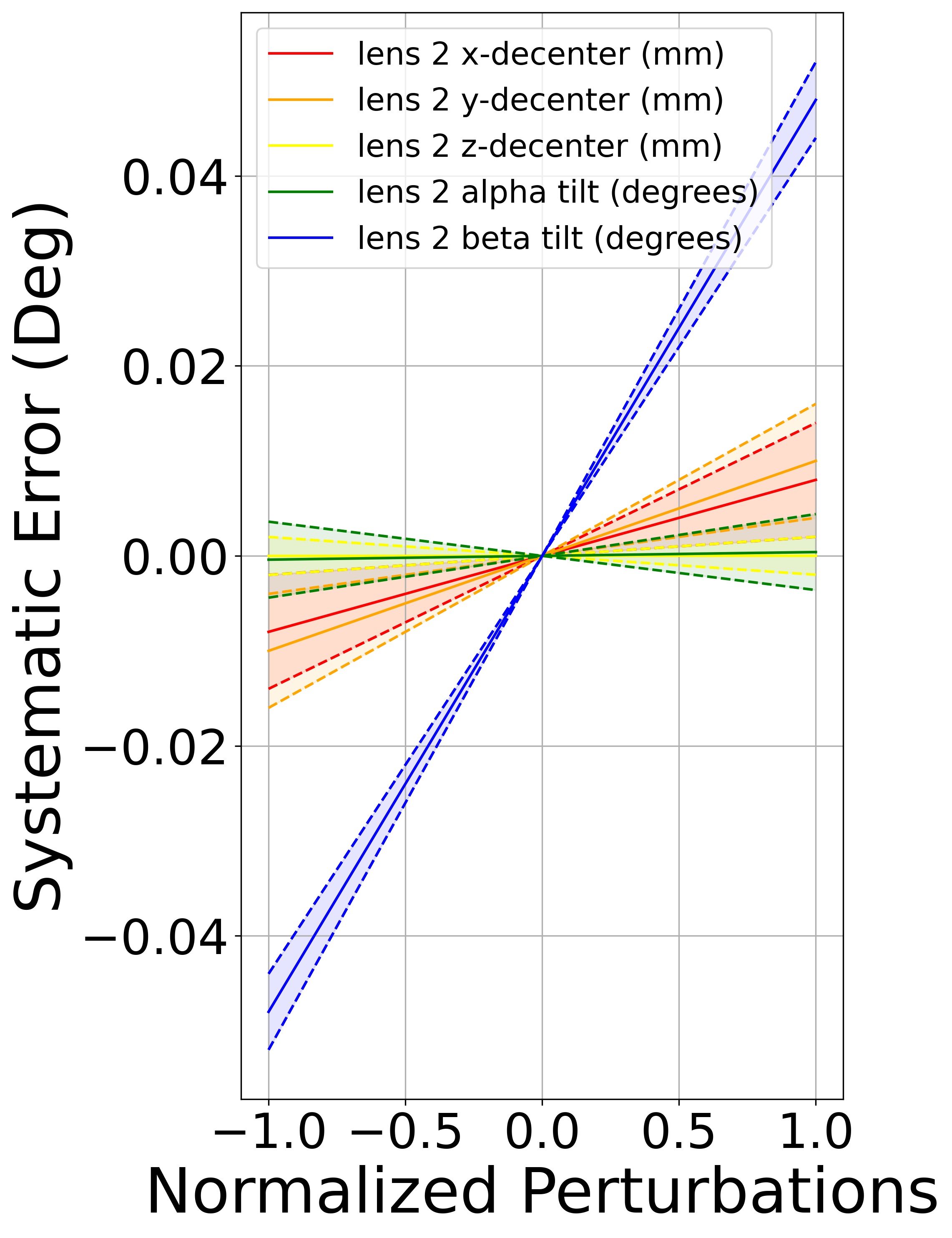}
        \vspace{-0.25in}
         \caption{PA6 Lens 2}
         \vspace{0.125in}

     \end{subfigure}

     \begin{subfigure}[b]{0.3\textwidth}
         \centering
         \includegraphics[width=\textwidth]{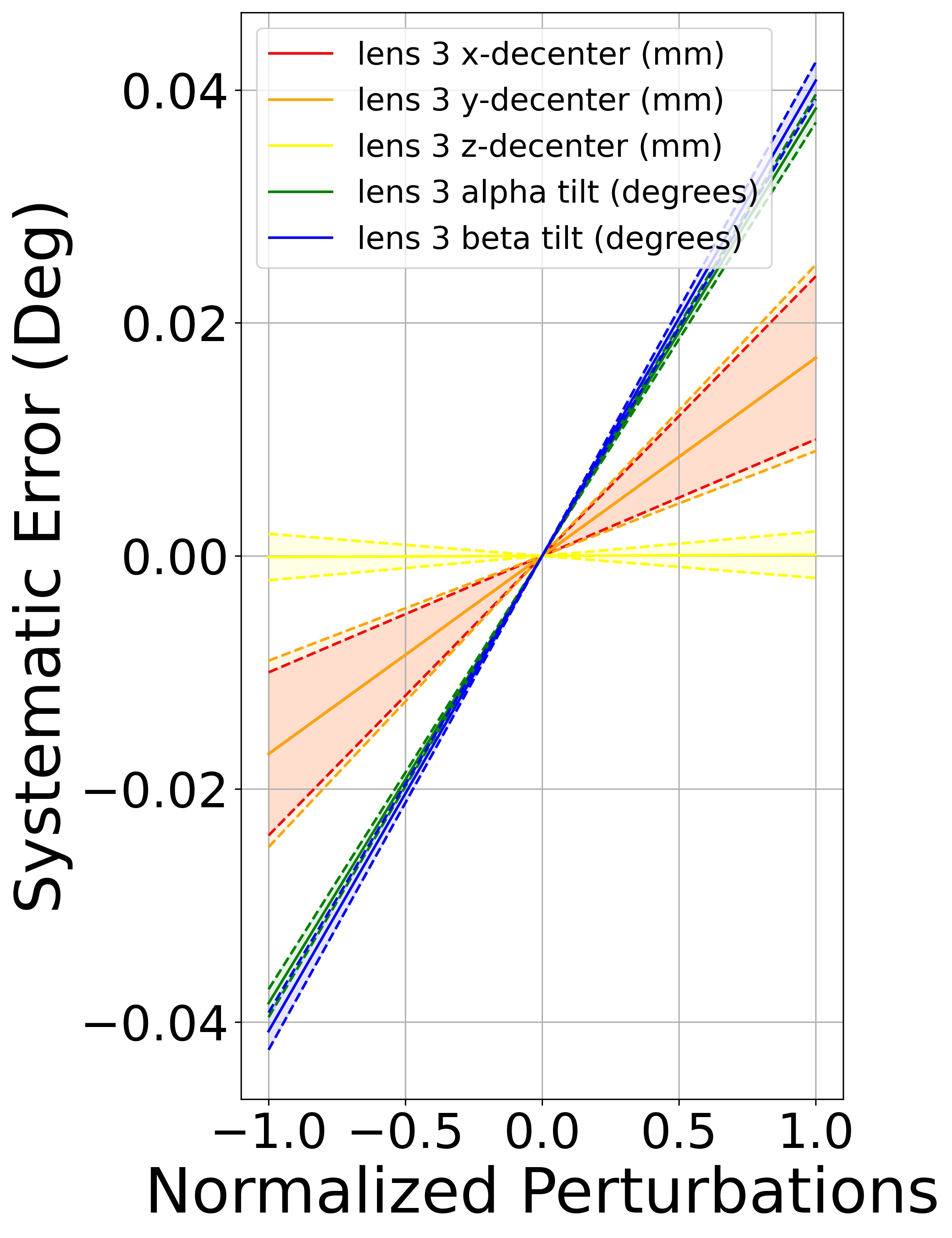}
        \vspace{-0.25in}
         \caption{PA4 Lens 3}
     \end{subfigure}
     \begin{subfigure}[b]{0.3\textwidth}
         \centering
         \includegraphics[width=\textwidth]{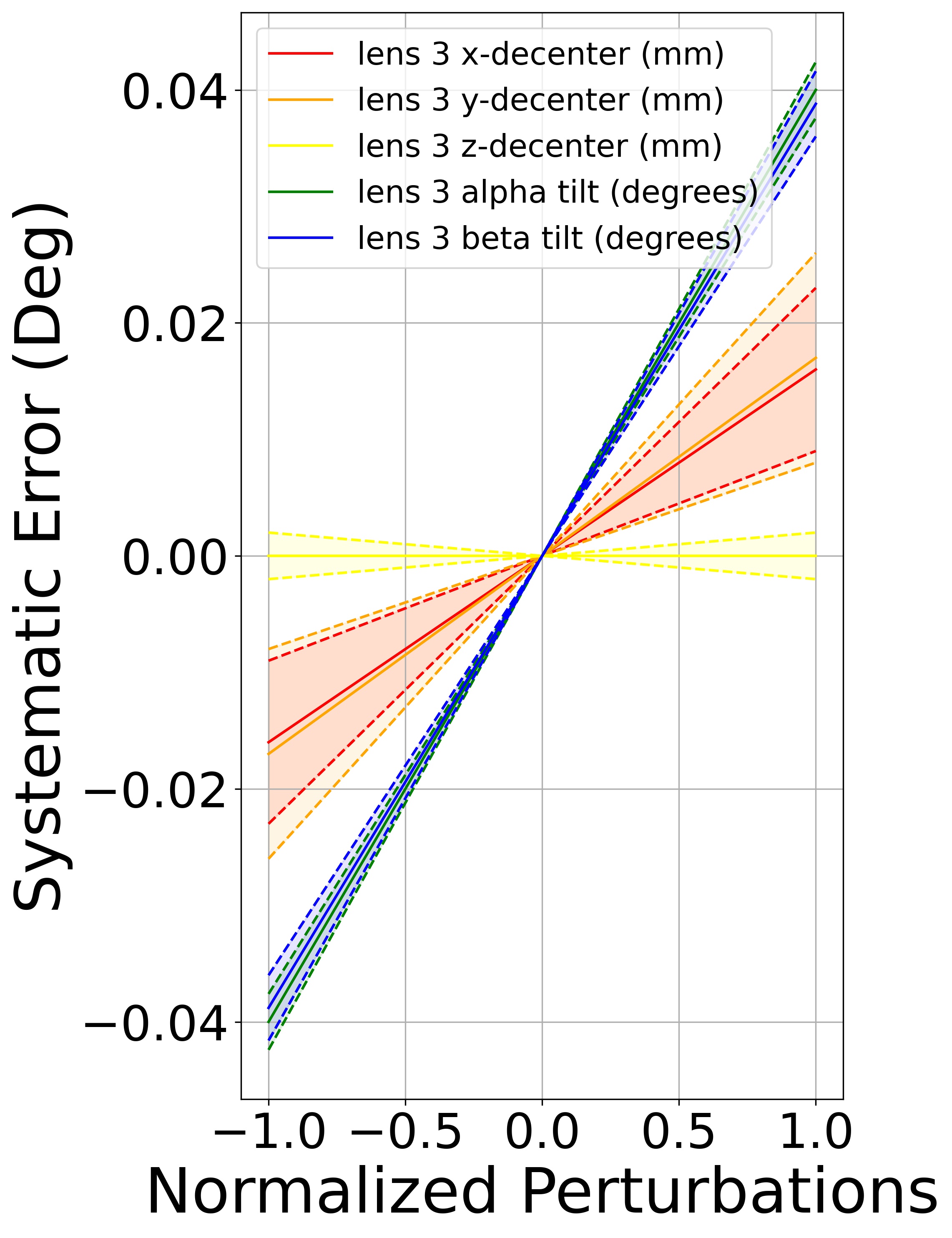}
        \vspace{-0.25in}
         \caption{PA5 Lens 3}

     \end{subfigure}
     \begin{subfigure}[b]{0.3\textwidth}
         \centering
         \includegraphics[width=\textwidth]{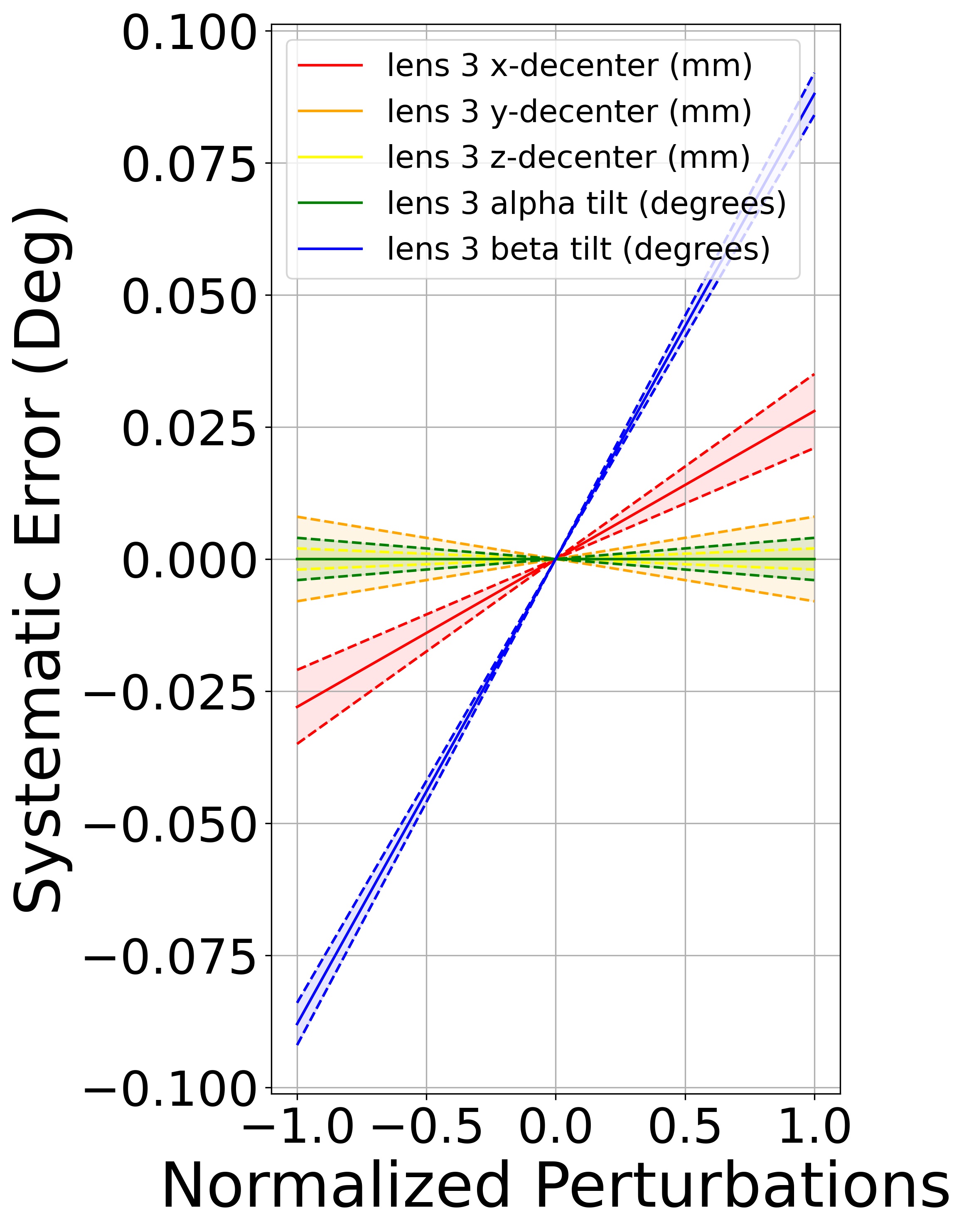}
        \vspace{-0.25in}
         \caption{PA6 Lens 3}
     \end{subfigure}
        \caption{Systematic uncertainty from perturbing each lens in PA4--6 (continued from Fig.~\ref{fig:FOV-average}).}
        \label{fig:summary_figures}
\vspace{-0.2in}
\end{figure}

Systematic uncertainties in the instrument polarization angles as a result of perturbations in the $x$, $y$, $z$, $\alpha$ and $\beta$ positions of the lenses and optics tubes are calculated using the above methodology for PA4--6. As mentioned in Section~\ref{subsec:syst_uncert}, systematic uncertainties due to perturbations of the ARC thickness and index of refraction for each array are found to be negligible relative to perturbations in $x$, $y$, $z$, $\alpha$ and $\beta$ (see Appendix~\ref{appen:individual_perturbation} and Figures~\ref{fig:FOV-average} and ~\ref{fig:summary_figures}). A typical example of the variation of the polarization angle across a detector plane is shown in Fig.~\ref{fig:pa3 lens 1 beta tilt} for a 1$^{\circ}$ tilt in $\beta$ for the optics tube of PA4. For each lens and optics tube, the systematic uncertainties for each perturbed optical parameter are plotted against their normalized perturbation values in Figs.~\ref{fig:FOV-average} and ~\ref{fig:summary_figures} for ease of comparison. The overall systematic uncertainties for PA4--6 are given in Table~\ref{tab:overall}. The specific results for individual perturbations are presented in Appendix~\ref{appen:macro}.

\begin{table}[b!]

\caption{Overall instrument polarization angle calibration and systematic uncertainties. The second column shows the range of instrument polarization for the unperturbed optical model (from Fig.~\ref{fig:advact_det_sky_coords}). The third column gives the systematic uncertainty estimates based on equations 3 and 4 ($\delta \pm \sigma$).} 
\centering 
\begin{tabular}{c c c c} 
\hline\hline 
Array & Instrument polarization corrections ($\deg$) \cite{Koopman} & Systematic uncertainty ($\deg$) \\ [0.5ex] 
\hline 
PA4 & [1.62, 0.51] & $0.09 \pm 0.03 $\\
PA5 & [$-$1.63, $-$0.45] & $0.09 \pm 0.04$\\
PA6 & [$-$0.35, 0.35] & $0.11 \pm 0.03$\\
[1ex] 
\hline 
\label{tab:overall}
\end{tabular}
\vspace{-0.2in}
\end{table}

Given our baseline assumptions for the level of optical parameter perturbations, we find the overall systematic uncertainties for each of the arrays to be comparable to the DR4 sensitivity \cite{Choi_2020}. As already done with DR4, assessing the internal consistency of $\psi$ measured from largely heterogeneous datasets for DR6 (from different detector arrays and observing seasons) could further reduce the systematic uncertainties estimated here. Additionally, more precise knowledge about the actual locations and orientations of the lenses and elements beyond our baseline assumptions would improve the systematic uncertainties for the $\psi$ constraints with DR6 and beyond. This would also provide corrections to the existing calibrations from \cite{Koopman}. Such information, in principle, could be obtained by opening up the AdvACT receiver and measuring the optical elements in the receiver. For each array, the largest contributor to the systematic uncertainty is a tilt in $\beta$ for the third lens, followed by a tilt in $\beta$ for the optics tube.  

\section{Discussion}\label{conclusion}
The absolute detector polarization angles for ACT are determined sufficiently through the clocking of arrays based on pointing measurements, in combination with position-dependent instrument polarization corrections derived from optical modeling \cite{Choi_2020,Koopman_2016}. In this paper, we used polarization-sensitive ray tracing in CODE V to estimate the overall systematic uncertainties associated with these instrument polarization corrections for the AdvACT arrays, ranging from 0.09$^{\circ}$ to 0.11$^{\circ}$. The systematic uncertainties presented here are expected to be larger than the statistical uncertainties in DR6. We note assessing internal consistency of independent datasets from different arrays and observing seasons could empirically limit (indicate) the actual systematic uncertainty to be below (above) our estimates. For instance, the ACTPol arrays (PA1–3), which utilized equivalent optics to AdvACT, exhibited good consistency in the EB null angle ($\psi$) estimates within their statistical uncertainties (typically ranging between 0.2$^{\circ}$ and 0.7$^{\circ}$) \cite{Choi_2020}. 

Previous measurements of the ACT relative detector polarization angles using wire grids are consistent with the results obtained from optical modeling \cite{Koopman_2016, Koopman}. For CCAT, more precise measurements using higher quality wire grids are planned to further validate the optical modeling of the refractive optics. Additionally, employing external calibrators for absolute polarization angles, such as polarization sources mounted on drones or CubeSats \cite{Coppi_2022, Johnson_2015}, could offer important redundancy in polarization angle calibration for upcoming observatories. Recent development with holography-based measurements also provides a complementary means of characterizing the polarization properties of the optics tubes \cite{Sierra_2024}.

\begin{figure}
     \centering
     \begin{subfigure}[b]{0.45\textwidth}
         \centering
         \includegraphics[width=\textwidth]{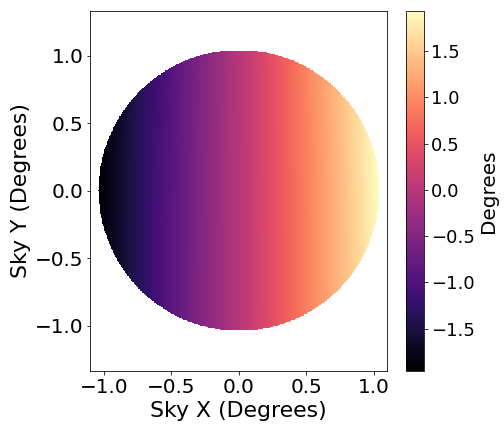}
         \caption{ACT (Gregorian  Mirrors)}
         \label{fig:ACT mirros}
     \end{subfigure}
     \begin{subfigure}[b]{0.45\textwidth}
         \centering
         \includegraphics[width=\textwidth]{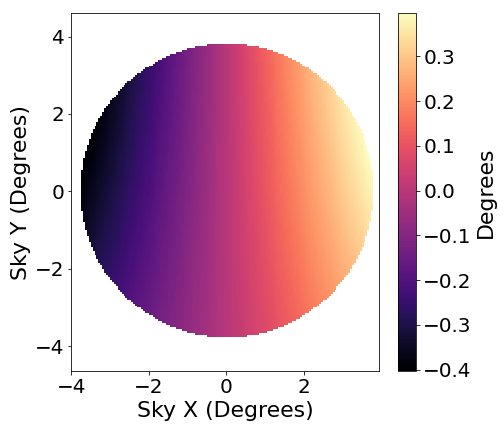}
         \caption{CCAT/SO (Crossed-Dragone Mirrors)
}
         \label{fig:CD mirrors}
    \end{subfigure}
    \caption{FOV maps of the telescope-only instrument polarization for ACT, which features an aplanatic Gregorian mirror design, and CCAT/SO \cite{Parshley_2018}, which features a Crossed Dragone mirror design. These preliminary results suggest that the CCAT/SO telescope optics have a $\sim$5$\times$ smaller polarization rotation than the ACT telescope optics. These calculations do not include the receiver.}
    \label{telescope_only}
\vspace{-0.2in}
\end{figure}

The method of polarization-sensitive ray tracing in CODE V demonstrated here and in \cite{Koopman} can be used to calculate the nominal instrument polarization angles and their systematic uncertainties for next-generation ground-based CMB observatories, including the Simons Observatory (SO), CCAT's Fred Young Submillimeter Telescope, and CMB-S4 \cite{SO/2019,CCAT/2023,S4/2016}. As shown in Fig. 3.8 in \cite{Koopman}, the upper bound for the in-band instrument polarization is set by the polarization rotations due to the telescope optics. Instrument polarization due to the telescope optics alone (Fig.~\ref{telescope_only}) for ACT and the Crossed-Dragone mirror design used for CCAT and SO suggest that instrument polarization rotations will be a $\sim$5$\times$ smaller effect for these observatories \cite{Niemack:16, Parshley_2018}. Our perturbative simulation based method combined with other metrology and data-driven tools can provide a powerful handle on understanding polarization angle measurement limits for future microwave observatories.

In future ACT measurements if either the EB signal is detected to be non-zero or inconsistencies are found between arrays, it would be necessary to check for instrumental effects that are not considered in our current analysis.  Some candidates include imperfections in optical elements (windows, filters, lenses, and anti-reflection coatings as discussed in Section~\ref{subsec:syst_uncert}), detector antenna cross-polarization leakage, and cross-talk between detectors that mixes polarization states. As described above, published analyses of ACT data thus far have been consistent with our calculated polarization rotations and systematic uncertainties presented here. If future measurements detect polarization rotations that are inconsistent with this approach, that could motivate additional studies and laboratory measurements of alternative sources of polarization rotation. 
\nocite{Zhang_17}

\begin{backmatter}
\bmsection{Acknowledgments}
This work was supported by the U.S. National Science Foundation through awards AST-0408698, AST-0965625, and AST-1440226 for the ACT project, as well as awards PHY-0355328, PHY-0855887 and PHY-1214379. Funding was also provided by Princeton University, the University of Pennsylvania, and a Canada Foundation for Innovation (CFI) award to UBC. ACT operates in the Parque Astron\'omico Atacama in northern Chile under the auspices of the Comisi\'on Nacional de Investigaci\'on (CONICYT). We thank the ACT collaborators and Eiichiro Komatsu for helpful comments. SKC acknowledges support from NSF award AST-2001866. MDN acknowledges support from NSF award AST-2117631. 

\bmsection{Disclosures}
The authors declare no conflicts of interest.

\end{backmatter}

\bibliography{sample}

\appendix

\section{Poldsp Macro Comparison to 3D Coherence Matrix Model for Polarization-Sensitive Ray Tracing}
\label{appen:macro}
\texttt{Poldsp} is a 2-dimensional polarization-sensitive ray tracing algorithm compatible with CODE V. It assumes that light is always polarized perpendicularly to the direction of propagation. This assumption is valid in the limit of small entrance pupil diameters, narrow fields of view, and small numerical apertures. A 2017 article published in Optics Express compared \texttt{poldsp} calculations of the polarization state to a more robust 3D coherence matrix model for a system with an focal ratio of 3 \cite{Zhang_17}. They found that the \texttt{poldsp} numbers agreed with the more sophisticated model up to about 8\%. 

\section{Individual Perturbation Results for PA 4--6}
\label{appen:individual_perturbation}
\begin{table}[ht]
\caption{Individual Perturbation Results for Arrays 4--6 (for Figures~\ref{fig:FOV-average} and ~\ref{fig:summary_figures})} 
\centering 
\resizebox{\textwidth}{!}{
\begin{tabular}{ l l | l l | l l | l l  } 
\hline\hline
 & & \multicolumn{2}{c}{PA4}& \multicolumn{2}{c}{PA5}& \multicolumn{2}{c}{PA6}\\
Perturbed Quantity & Units & Slope & Stdev & Slope & Stdev & Slope & Stdev\\ [0.5ex] 
\hline 
layer 1 arc thickness (\%) & degrees/arc thickness \% & 0.000 & 0.003 & 0.00 & 0.01 & 0.000 & 0.004 \\
layer 2 arc thickness (\%) & degrees/arc thickness \% & 0.000 & 0.003 & 0.000 & 0.009 & 0.000 & 0.003 \\
layer 3 arc thickness (\%) & degrees/arc thickness \% & 0.000 & 0.007 & 0.000 & 0.005 & 0.000 & 0.002 \\
layer 1 arc index (\%) & degrees/arc index \% & 0.000 & 0.005 & 0.0001 & 0.0002 & 0.000 & 0.0002 \\
layer 2 arc index (\%) & degrees/arc index \% & 0.0000 & 0.0003 & 0.0000 & 0.0001 & 0.0000 & 0.0002 \\
layer 3 arc index (\%) & degrees/arc index \% & 0.0000 & 0.0001 & 0.0000 & 0.0002 & 0.0000 & 0.0002 \\
optics tube x-decenter (mm) & degrees/mm decenter & 0.009 & 0.001 & 0.009 & 0.001 & 0.0 & 0.001 \\
optics tube y-decenter (mm) & degrees/mm decenter & 0.001 & 0.001 & 0.001 & 0.001 & 0.0 & 0.00055\\
optics tube z-decenter (mm) & degrees/mm decenter & 0.0005 & 0.0004 & 0.0006 & 0.0004 & 0.0 & 0.0 \\
optics tube alpha tilt (degrees) & degrees/degrees alpha tilt & 0.016 & 0.012 & 0.016 & 0.012 & 0.0 & 0.012 \\
optics tube beta tilt (degrees) & degrees/degrees beta tilt & 0.044 & 0.012 & 0.044 & 0.012 & 0.048 & 0.012 \\
lens 1 x-decenter (mm) & degrees/mm decenter & 0.017 & 0.003 & 0.017 & 0.003 & 0.039 & 0.004 \\
lens 1 y-decenter (mm) & degrees/mm decenter & 0.016 & 0.004 & 0.016 & 0.004 & 0.0 & 0.005 \\
lens 1 z-decenter (mm) & degrees/mm decenter & 0.0001 & 0.002 & 0.0 & 0.002 & 0.0 & 0.003 \\
lens 1 alpha tilt (degrees) & degrees/degrees alpha tilt & 0.0 & 0.016 & 0.0 & 0.016 & 0.0 & 0.016 \\
lens 1 beta tilt (degrees) & degrees/degrees beta tilt & 0.0028 & 0.0144 & 0.004 & 0.016 & 0.004 & 0.016 \\
lens 2 x-decenter (mm) & degrees/mm decenter & 0.008 & 0.006 & 0.007 & 0.006 & 0.016 & 0.005 \\
lens 2 y-decenter (mm) & degrees/mm decenter & 0.01 & 0.006 & 0.01 & 0.007 & 0.0 & 0.006 \\
lens 2 z-decenter (mm) & degrees/mm decenter & 1e-05 & 0.002 & 0.0 & 0.002 & 0.0 & 0.006 \\
lens 2 alpha tilt (degrees) & degrees/degrees alpha tilt & 0.0004 & 0.004 & 0.0004 & 0.0032 & 0.0 & 0.008 \\
lens 2 beta tilt (degrees) & degrees/degrees beta tilt & 0.048 & 0.004 & 0.0476 & 0.0048 & 0.012 & 0.004 \\
lens 3 x-decenter (mm) & degrees/mm decenter & 0.017 & 0.007 & 0.016 & 0.007 & 0.028 & 0.007 \\
lens 3 y-decenter (mm) & degrees/mm decenter & 0.017 & 0.008 & 0.017 & 0.009 & 0.0 & 0.008 \\
lens 3 z-decenter (mm) & degrees/mm decenter & 0.0001 & 0.002 & 0.0 & 0.002 & 0.0 & 0.002 \\
lens 3 alpha tilt (degrees) & degrees/degrees alpha tilt & 0.038 & 0.0012 & 0.04 & 0.0024 & 0.0 & 0.004 \\
lens 3 beta tilt (degrees) & degrees/degrees beta tilt & 0.0408 & 0.0016 & 0.0388 & 0.0028 & 0.088 & 0.004 \\
[1ex] 
\hline 
\end{tabular}
}
\label{table:nonlin} 
\end{table}

\end{document}